\documentclass[prd,twocolumn,showpacs,preprintnumbers,floats,floatfix,
amssymb,amsmath,nofootinbib]{revtex4}
\usepackage{graphicx}
\usepackage{dcolumn}
\usepackage{bm}
\def\spose#1{\hbox to 0pt{#1\hss}}
\def\lta{\mathrel{\spose{\lower 3pt\hbox{$\mathchar"218$}}
     \raise 2.0pt\hbox{$\mathchar"13C$}}}
\def\gta{\mathrel{\spose{\lower 3pt\hbox{$\mathchar"218$}}
     \raise 2.0pt\hbox{$\mathchar"13E$}}}
\newcommand{\be}{\begin{equation}}
\newcommand{\ee}{\end{equation}}
\newcommand{\bea}{\begin{eqnarray}}
\newcommand{\eea}{\end{eqnarray}}

\begin{document}


\title{On Signatures of Short Distance Physics in the Cosmic
Microwave Background} 

\author{Robert~H.~Brandenberger}
\email{rhb@het.brown.edu}
\address{Theory Division, CERN, CH-1211 Geneva 23, Switzerland 
\\
and
\\
Department of Physics, Brown University, Providence, RI 02912, USA} 

\author{J\'er\^ome Martin} 
\email{jmartin@iap.fr}
\address{Institut d'Astrophysique de Paris, 98bis  boulevard Arago,
75014 Paris, France} 
\date{\today}


\begin{abstract}
Following a self-contained review of the basics of the theory of
cosmological perturbations, we discuss why the conclusions reached in
the recent paper by Kaloper et al.~\cite{Kaloper} are too pessimistic
estimates of the amplitude of possible imprints of trans-Planckian
(string) physics on the spectrum of cosmic microwave anisotropies in
an inflationary Universe. It is shown that the likely origin of large
trans-Planckian effects on late time cosmological fluctuations comes
from nonadiabatic evolution of the state of fluctuations while the
wavelength is smaller than the Planck (string) scale, resulting in an
excited state at the time that the wavelength crosses the Hubble
radius during inflation.
\end{abstract}


\maketitle

\section{Introduction}

\begin{widetext}

\begin{figure}[htbp]
\includegraphics*[width=18cm, height=10cm, angle=0]{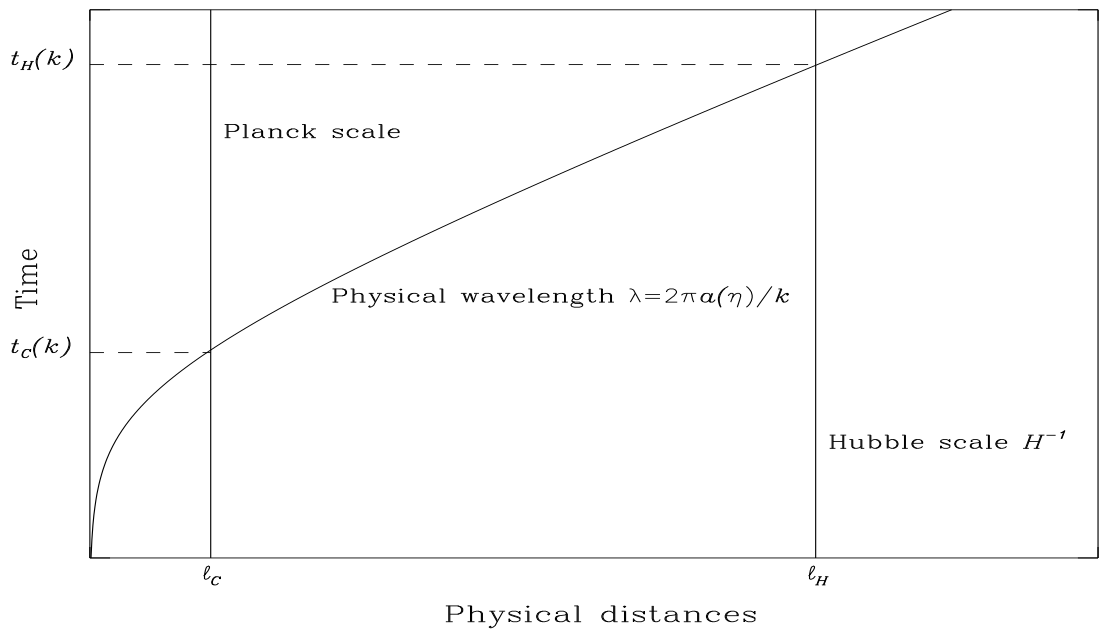}
\caption{Space-time sketch of the evolution of
a comoving length scale with comoving wavenumber $k$
in an inflationary Universe. The coordinates are
physical distance and cosmic time $t$. At very
early times, the wavelength is smaller than the
Planck scale $\ell _{\rm Pl}$ (Phase I), at intermediate times
it is larger than $\ell _{\rm Pl}$ but smaller than the Hubble
radius $H^{-1}$ (Phase II), and at late times during
inflation it is larger than the Hubble radius (Phase III).}
\label{sketch}
\end{figure}
\end{widetext}

It has recently been emphasized that the predictions of inflationary
cosmology for the spectrum of density fluctuations and Cosmic
Microwave Background (CMB) anisotropies may not be robust against
effects of trans-Planckian (string) physics~\cite{RHBrev}. This {\it
trans-Planckian problem} can easily be seen from the space-time sketch
in Fig.~\ref{sketch}. Essentially all current realizations of the
inflationary scenario are based on weakly interacting fields, in which
context the Fourier modes of the field representing cosmological
fluctuations evolve independently from the initial time (e.g. the
beginning of inflation) until their amplitude reaches order $1$ in the
recent past. Most models also have a period of inflation greatly in
excess of the minimal number required to solve the cosmological
problems of standard cosmology~\cite{inf,Linde}. Provided that the period
of inflation lasts more than about 70 e-foldings, then the physical
wavelength of comoving scales responsible for present CMB anisotropies
was smaller than the Planck (string) scale at the beginning of
inflation.  Hence, to study the evolution of fluctuations from the
time they are formed until the time their wavelength becomes larger
than the Planck (string) scale, the effects of trans-Planckian
(string) physics cannot be neglected.

The possible effect of trans-Planckian physics on the spectrum of
cosmological perturbations was first studied in detail in
Ref.~\cite{MB} (see also Ref.~\cite{Niemeyer}) by means of replacing
the standard free field theory dispersion relation by some ad hoc
dispersion relations. The same method and dispersion relations were
used in \cite{Unruh} and \cite{CJ} in the context of an analysis of
possible trans-Planckian effects on black hole radiation. It was found
that if the evolution of the modes is non-adiabatic in the initial
stages, then significant effects on the spectrum of cosmological
fluctuations are possible~\cite{MB3,NP2}. Subsequently, the
possibility of measurable effects of trans-Planckian physics on
observables such as CMB anisotropies and power spectra of scalar and
tensor metric fluctuations was studied
\cite{kempf,Chu,Easther,kempfN,Hui,HoB} in models where the trans-Planckian
physics is based on stringy space-time uncertainty relations, and in
some examples large effects were found.

Very recently, a paper has appeared~\cite{Kaloper} which claims to
show using general effective field theory techniques that
trans-Planckian effects on CMB anisotropies in an inflationary
Universe must be suppressed by a factor of $(H_{\rm inf}/M)^2$, where
$H_{\rm inf}$ is the Hubble constant during inflation, and $M$ is the
scale of trans-Planckian physics. This result implies that
trans-Planckian effects are not observable. This conclusion appears to
be in conflict with the analyses of Refs.~\cite{Easther,Hui,HoB}.

The purpose of this note is to point out that the conclusions of
Ref.~\cite{Kaloper} are too pessimistic concerning the potential
observability of trans-Planckian (string) physics in the spectrum of
CMB anisotropies. The key point is that in Ref.~\cite{Kaloper}, the
effect of trans-Planckian physics on the amplitude of
fluctuations of a particular Fourier mode of the fluctuating field is
estimated at the time the mode crosses the Hubble radius during
inflation, and assuming that the state of this mode is the local vacuum
state at that time. Recall that in an expanding background, the vacuum
state of a scalar field on this background - and the fields which
characterize cosmological perturbations are such scalar fields - is
not uniquely defined. In particular, a state which at early times is
empty of particles in the comoving frame will in general appear to
contain many particles at a later time \cite{Parker,BDbook}. The
effect described in Ref.~\cite{Kaloper} is indeed usually very
small. However, the more important effect of trans-Planckian physics
is to open the possibility of a non-adiabatic evolution of the initial
local vacuum (the local vacuum at the initial time, e.g. the beginning
of inflation) on trans-Planckian scales, thus leading to a state of
the fluctuation mode at the time of Hubble radius crossing which is
highly excited~\cite{MB}, or to lead to other effects which can be
characterized as changing the initial conditions on the state at the
time when the fluctuation mode crosses the Hubble radius.

To make these points clear, it is important to dispel the myth that
in current models of inflation, based on weakly coupled fields, the
fluctuations are generated at the time they cross the Hubble
radius. Thus, in the following section we will give an overview of the
quantum theory of the generation and evolution of cosmological
fluctuations, hopefully providing a pedagogical introduction to this
subject. In Sec.~III, we then compare the analysis of
\cite{Kaloper} with the studies which have shown that trans-Planckian
physics may leave imprints in physical quantities such as CMB anisotropies
which are observable. Finally, in Sec.~IV, we present our conclusions.

\section{Theory of Cosmological Perturbations}

In the following we give an overview of the quantum theory of
cosmological perturbations. The reader is referred to \cite{MFB} for
details and references to the original literature (see for example
Ref.~\cite{MS}. A modern textbook treatment can also be found in
\cite{LLbook}. Since gravity is a purely attractive force, and since
the fluctuations on scales of the CMB anisotropies were small when the
anisotropies were generated, the fluctuations had to have been very
small in the early Universe. Thus, a linearized analysis of the
fluctuations is justified. In this case, the Fourier modes of the
cosmological fluctuations evolve independently.  
\par

The basic idea of the theory of cosmological perturbations (which
includes the theory of gravitational waves) is to quantize the linear
fluctuations about a classical background cosmology described by a
homogeneous and isotropic Friedmann cosmology with metric
\begin{equation}
\label{FRW} 
{\rm d}s^2 \, = \, a^2(\eta)\bigl( {\rm d}\eta^2 - {\rm
d}{\bf x}^2 \bigr). 
\end{equation} 
Here, $\eta$ is conformal time related to the physical time $t$ via
${\rm d}t = a(\eta ) {\rm d}\eta$, and we have considered for
simplicity the case of a spatially flat Universe. The coordinates
${\bf x}$ are comoving coordinates. The starting point is the full
action of gravity plus matter 
\begin{equation}
S \, = \, \int {\rm d}^4x \sqrt{-g}R +S_{\rm m} , 
\end{equation}
where the first term is the usual Einstein-Hilbert action for gravity,
$R$ being the Ricci scalar and $g$ the determinant of the metric, and
$S_{\rm m}$ is the matter action. For the sake of simplicity, and
since it is the usual assumption in simple inflationary Universe
models, we take matter to be described by a single minimally coupled
scalar field $\varphi$. Then, we separate the metric and matter into
classical background variables $g^{(0)}_{\mu \nu}, \varphi^{(0)}$
which depend only on time, and fluctuating fields $h_{\mu \nu},
\delta\varphi$ which depend on space and time and have vanishing
spatial average:
\begin{equation}
g_{\mu \nu} \, =\, g^{(0)}_{\mu \nu}(\eta) +
h_{\mu \nu}(\eta ,{\bf x}), 
\quad 
\varphi \, =\,\varphi^{(0)}(\eta ) +
\delta\varphi (\eta ,{\bf x})\, .  
\end{equation}
There are two kinds of metric perturbations of interest in early
Universe cosmology: the scalar and tensor fluctuations
\footnote{Vector fluctuations are redshifted in expanding cosmological
backgrounds and hence are not usually considered.}. At the level of
the linearized equations of motion there is no coupling between scalar and
tensor modes, and thus they can be quantized independently. 
\par

Let us first consider tensor fluctuations. Tensor fluctuations
correspond to gravitational waves. The perturbed metric only has
non-vanishing space-space components $h_{i j}(\eta ,{\bf x})$, 
\begin{equation}
{\rm d}s^2=a^2(\eta )[{\rm d}\eta ^2-(\delta _{ij}+h_{ij}){\rm d}x^i
{\rm d}x^j],
\end{equation}
which can be expanded in terms of the two basic traceless and
symmetric polarization tensors $e^{+}_{i j}$ and $e^{\times}_{i j}$ as
\begin{equation}
h_{i j}(\eta ,{\bf x}) \, = \, h_{+}e^{+}_{i j} + 
h_{\times }e^{\times }_{i j} \,
\end{equation}
where the space and time dependence is in the coefficient functions
$h_{+}$ and $h_{\times }$. When the Einstein action is expanded to
second order in the metric fluctuations about a
Friedmann-Robertson-Walker (FRW) background~(\ref{FRW}), the action
for $h_{+}$ and $h_{\times }$ reduces to that of a free, massless,
minimally coupled scalar field $h$ in the FRW background.  To obtain
the correct normalization, the metric must be multiplied by the
normalization factor $m_{\rm Pl}/\sqrt{2}$, where $m_{\rm Pl}$
is the four-dimensional Planck mass. In order to obtain the
equation of motion, we expand the action to second order in the
fluctuating fields (the terms in the action linear in the fluctuating
fields vanish if the background is taken to be a solution of the
equations of motion). In Fourier space, the action is
\begin{equation}
\label{gravac}
\delta S_{\rm g}=\int {\rm d}\eta \frac{a^2}{2} 
\biggl( h'_{-{\bf k}}h'_{\bf k} -
k^2h_{-{\bf k}}h_{\bf k}\biggr) \, .  
\end{equation}
This leads to the equation of motion 
\begin{equation}h_{\bf k}''+ 2 {{a'} \over a} h_{\bf k}'+ k^2 h_{\bf k} \, 
= \, 0 \, .
\end{equation}
The Hubble friction term can be eliminated via a change of variables
$\mu_k \, \equiv \, a h_k \, $, yielding the equation of motion
\begin{equation}\label{graveq}
\mu_{\bf k}''+ \biggl( k^2 - {{a''} \over a} \biggr) \mu_{\bf k} \, 
= \, 0 \, .
\end{equation}
One recognizes the equation of motion of a parametric oscillator, 
an oscillator with a time-dependent fundamental frequency.
\par

Let us now turn the second type of cosmological perturbations: scalar
perturbations. Scalar metric fluctuations couple to matter, and give
rise to the large-scale structure of the Universe. The description of
scalar metric perturbations is more complicated than the analysis of
gravitational waves both because of the coupling to matter and also
because some perturbation modes correspond to space-time
reparametrizations of a homogeneous and isotropic cosmology. This is
the issue of gauge fixing.  A simple way to address this issue is to
work in a system of coordinates which completely fixes the gauge. A
simple choice is the {\it longitudinal} gauge, in which the metric takes
the form \cite{MFB}
\begin{equation}
{\rm d}s^2 \, = \, a^2(\eta) \bigl[(1 + 2 \Phi) d\eta^2 - 
(1 - 2 \Psi) \delta_{i j}
{\rm d}x^i {\rm d}x^j \bigr] \, ,
\end{equation}
where the space and time dependent functions $\Phi$ and $\Psi$ are the
two physical metric degrees of freedom which describe scalar metric
fluctuations The fluctuations of matter fields give additional degrees
of freedom for scalar metric fluctuations. In the simple case of a
single scalar matter field, the matter field fluctuation can be
denoted by $\delta \varphi$.  In the absence of anisotropic stress, it
follows from the Einstein equations that the two metric fluctuation
variables $\Phi$ and $\Psi$ coincide. Due to the Einstein constraint
equation, the remaining metric fluctuation $\Psi$ is determined by the
matter fluctuation $\delta \varphi$. It is clear from this analysis of
the physical degrees of freedom that the action for scalar metric
fluctuations must be expressible in terms of the action of a single
free scalar field $v$ with a time dependent mass (determined by the
background cosmology). As shown in \cite{M85} (see also \cite{L80}),
this field is
\begin{equation}
v \, =\, a \biggl( \delta \varphi + {{\varphi_0'} 
\over {\cal H}} \Psi \biggr) =z {\cal R} \, ,
\end{equation}
where $\varphi_0(\eta )\equiv \varphi ^{(0)}$ denotes the background 
value of the scalar matter field, 
${\cal H} = a'/ a$, 
\begin{equation}\label{zdef}
z \, \equiv \, a {{\varphi_0'} \over {\cal H}} \, ,
\end{equation}
and ${\cal R}$ denotes the curvature perturbation in comoving 
gauge \cite{L85}. The action for scalar metric fluctuations is 
\cite{M88}
\begin{equation}\label{scaleac}
\delta S_{\cal R}\, =\, {1 \over 2} \int {\rm d}^4{\bf x} 
\biggl[(v_{\bf k}')^2- \delta^{i j} 
v_{{\bf k}, i} v_{{\bf k}, j} + {{z''} \over z} v_{\bf k}^2 
\biggr] \, ,
\end{equation}
which leads to the equation of motion
\begin{equation}\label{scaleeq}
v_{\bf k}'' + \biggl( k^2 - {{z''} \over z} \biggr) v_{\bf k} \, 
= \, 0 \, ,
\end{equation}
which under the change $a \rightarrow z$ is identical to the
equation~(\ref{graveq}) for gravitational waves. Therefore, we obtain
again the equation of a parametric oscillator. Note that if $a(\eta)$
is a power of $\eta$, then $\varphi_0'$ and ${\cal H}$ scale with the
same power of $\eta$. The variable $z$ is then proportional to $a$, and thus
the evolution of gravitational waves and scalar metric fluctuations is
identical. In this case, the solution can be expressed in terms of
Bessel functions.
\par

Let us now analyze the behavior of the classical mode functions
$\mu_{\bf k}(\eta)$ and $v_{\bf k}(\eta)$. The equations
(\ref{graveq}) and (\ref{scaleeq}) are harmonic oscillator equations
with a time-dependent mass given by $a''/a$ and/or $z''/z$. On scales
smaller than the Hubble radius, the mass term is negligible, and the
mode functions oscillate with constant amplitude. On scales larger
than the Hubble radius, however, the mass term dominates and the $k^2$
term can be neglected. The mode functions no longer oscillate. In an
expanding background, the dominant mode of $\mu_{\bf k}(\eta)$ and
$v_{\bf k}(\eta)$ scales as $a(\eta)$.  Thus, the role of the time of
Hubble radius crossing is to give the time when the classical mode
functions begin to increase in amplitude.
\par

So far, all the considerations are classical. This is sufficient to
describe the evolution of the perturbations. However, if one is
interested in the source of the fluctuations, then a quantum treatment
becomes necessary. In this framework, the state of each mode of the
fluctuating field is fixed at some initial time $t_{\rm i}$ which (at
least in the context of cosmology described by the above action) is
independent of $k$ and can be taken to be the beginning of the period
of inflation. Note that in the framework currently used in
inflationary cosmology it is wrong to consider that fluctuations on
scale $k$ are generated at the time $t_{\rm H}(k)$ when that scale crosses
the Hubble radius. 
\par

The quantum description can be discussed most easily in the Heisenberg
picture in which the states are time-independent but the operators
evolve. From the action (\ref{scaleac}) it follows that the momentum
canonically conjugate to the field $v$ is $\Pi_{\bf k} \, = \, 
v'_{-{\bf k}} \,$ and this leads to the Hamiltonian 
\begin{equation}
\label{H} H_{\cal R} \, = \, \int
{\rm d}^3{\bf x} \biggl( \Pi^2_{\bf k} + \delta^{i j}v_{{\bf k}, i} v_{{\bf
k}, j} - {{z''} \over z}v^2_{\bf k} \biggr), 
\end{equation}
We now canonically quantize this Hamiltonian, elevating $v$ and $\Pi$
to canonically conjugate operators ${\hat v}$ and ${\hat \Pi}$, and
imposing the Hamilton equations as equations of motion. It then
immediately follows that the operator ${\hat v}$ satisfies the same
equation of motion as the classical field $v$. We can expand the
operator ${\hat v}$ into a basis of operators ${\hat c}_{\bf k}$
and ${\hat c}^{\dagger }_{\bf k}$ which, at the initial time $\eta_{\rm i}$, 
correspond to the Minkowski field creation and annihilation
operators. Specifically,
\begin{equation}
\label{candecomp}
{\hat v}(\eta ,{\bf x}) \, 
= \, \frac{1}{(2\pi )^{3/2}}\int {\rm d}^3{\bf k} 
[{\hat c}_{\bf k}(\eta )e^{i {\bf k} \cdot {\bf x}} 
+{\hat c}^{\dagger}_{\bf k}(\eta )e^{-i {\bf k} \cdot {\bf x}}
] \, , 
\end{equation}
The difference with the case of a free field is that, due to the
interaction of the field ${\hat v}(\eta ,{\bf x})$ with the classical
background, the time dependence of the creation and annihilation
operators is no longer given by $e^{\pm i\omega \eta }$.  This is a
manifestation of the fact that particles creation is now possible.
The operators ${\hat c}_{\bf k}(\eta )$ and ${\hat c}^{\dagger}_{\bf
k}(\eta )$ obey the usual creation and annihilation operator algebra
\begin{equation}
[{\hat c}_{\bf k}(\eta ), {\hat
c}^{\dagger }_{\bf p}(\eta )] \, = \, \delta({\bf k} - {\bf p})\, .
\end{equation}
This relation is of course valid for any time $\eta $. As initial
state, we choose the state which is empty of particles from the point
of view of the local comoving observer at the initial time $\eta _{\rm
i}$. This state $|0\rangle$ is defined by
\begin{equation}
{\hat c}_{\bf k}\vert 0\rangle \, = \, 0 \, .  
\end{equation}
Since due to the time dependence of the background there is nontrivial
mixing between creation and annihilation operators at different times,
this state is in general not the vacuum at later times. Since the mode
equation for fixed ${\bf k}$ has exactly two independent solutions,
the creation and annihilation operators at time $\eta >\eta _{\rm i}$
must be related to the creation and annihilation operators at initial
time $\eta _{\rm i}$ via a Bogoliubov transformation
\begin{eqnarray}
\label{bog1}
{\hat c}_{\bf k}(\eta )\, &=& \, 
\alpha _k (\eta ){\hat c}_{\bf k}(\eta _{\rm i})
+\beta _k(\eta )\hat{c}^{\dagger }_{-{\bf k}}(\eta
_{\rm i})\, ,
\\ 
\label{bog2}
{\hat c}_{\bf k}^{\dagger }(\eta )\, &=& \, 
\alpha _k ^*(\eta ){\hat c}_{\bf k}^{\dagger }(\eta _{\rm i})
+\beta _k^{*}(\eta )\hat{c}_{-{\bf k}}(\eta _{\rm i})\, ,
\end{eqnarray}
where the Bogoliubov coefficients $\alpha_k$ and
$\beta_k$ satisfy the normalization condition
\begin{equation}
|\alpha _k|^2 - |\beta _k|^2 \, = \, 1 \, .
\end{equation}
This relation guarantees that the commutation relations are preserved
in time.  The time dependence of the quantum field can now be written
as $\hat{c}_{\bf k}(\eta )=(\alpha _k+\beta _k^*)(\eta ) \hat{c}_{\bf
k}(\eta _{\rm i})\equiv v_k(\eta )\hat{c}_{\bf k}(\eta _{\rm i})$. The
temporal function $v_{\bf k}(\eta )$ is solution of the classical mode
equation and should be chosen to be pure positive frequency at the
time $\eta_{\rm i}$, with vacuum normalization
\begin{equation}
v_{\bf k}(\eta_{\rm i}) \, = \, {1 \over {\sqrt{2k}}} \, ,\quad v_{\bf
k}'(\eta_{\rm i}) \, = \, i \sqrt{\frac{k}{2}} \, .  
\end{equation}
This amounts to choosing $v_k^{\rm (in)}(\eta )\sim e^{-ik(\eta -\eta _{\rm
i})}/\sqrt{2k}$.  The Bogoliubov transformation exhibited in
Eqs.~(\ref{bog1}), (\ref{bog2}) gives the most general time evolution
of the quantum field ${\hat v}(\eta ,{\bf x})$. In particular, it is
not necessary that the ``out'' region be flat. If this is the case, 
then the most general mode function in the ``out'' region is a linear 
combination of positive and negative frequency plane waves, i.e., 
\begin{eqnarray}
v^{\rm (out)}_k(\eta ) &=& a_kv_k^{\rm (in)}(\eta )+b_k
v_k^{\rm (in)*}(\eta )\, ,
\\
v^{\rm (out)*}_k(\eta ) &=& a_k^*v_k^{\rm (in)*}(\eta )+b_k^*
v_k^{\rm (in)}(\eta )\, .
\end{eqnarray}
Inserting these relation into the canonical decomposition of the field
${\hat v}^{\rm (out)}(\eta ,{\bf x})$, see Eq.~(\ref{candecomp})
[using that $c_{\bf k}(\eta )=v_k(\eta )c_{\bf k}(\eta _{\rm i})$], and
comparing with the equation obtained using Eqs.~(\ref{bog1}),
(\ref{bog2}), one immediately reaches the conclusions that the 
coefficients $a_k$ and $b_k$ are in fact given by 
\begin{equation}
a_k=\alpha _k, \quad b_k=\beta _k^*.
\end{equation}
The ``number operator'' (using language appropriate to a
scalar field on a given background) which measures the number of 
particles of comoving momentum ${\bf k}$ from
the point of view of the comoving observer at time $\eta $
is
\begin{equation}
{\hat N}_{\bf k} (\eta )\, = \, {\hat c}_{\bf k}^{\dagger }(\eta )
{\hat c}_{\bf k}(\eta )\, .
\end{equation}
In the state $|0\rangle$, the expectation value of this number
operator is
\begin{equation}
\langle 0|{\hat N}_k|0\rangle \, = \, |\beta_k|^2 \, .
\end{equation}
The Bogoliubov coefficients $\beta_k$ thus measure the number of
particles from the point of view of the comoving observer at time
$\eta >\eta _{\rm i}$ in the state which at time $\eta_{\rm i}$ is the
local vacuum state.  Translated to field language, these Bogoliubov
coefficients measure the magnitude of the two point field correlation
function in momentum space at later times in the initial vacuum state.
Let us notice that the Bogoliubov coefficient can also be determined
by the overlap integral
\begin{equation}\label{scaleprod}
\beta_k \, = \, \langle v_k^{\rm (in)}, v_k^{\rm (out)*}\rangle  \, ,
\end{equation}
where $\langle a, b\rangle $ stands here for the usual Klein-Gordon
scalar product. 
\par

The quantum mechanical interpretation of the two phases $t<t_{\rm
H}(k)$ and $t>t_{\rm H}(k)$ is the following: on sub-Hubble scales we
have oscillating quantum vacuum fluctuations and the Bogoliubov
coefficients $\beta_k$ vanish. There is no particle production on
these scales.  Once the scales cross the Hubble radius, the mode
functions begin to grow and the fluctuations get frozen. By
(\ref{scaleprod}) this implies a growth of the Bogoliubov coefficients
$\beta_k$ which is proportional to the amplitude of $v_k(\eta)$.  The
initial vacuum state then becomes highly squeezed at $t\gg t_{\rm H}(k)$. In
the case of gravitational waves, this physics was first discussed in
\cite{Grishchuk}.  For a free scalar field on a cosmological
background, the squeezing of the initial quantum vacuum state
corresponds to particle production \cite{BDbook}.  Applied to the
fields representing cosmological fluctuations, the squeezing leads to
the generation of effectively classical cosmological perturbations.
\par

For cosmological applications, it is particularly interesting to
calculate the two-point correlation functions of gravitational waves
and density perturbations. For gravitational waves, the power spectrum
of gravitational waves in the vacuum state $|0\rangle $ can be written
in terms of the new field $\mu _k$ as
\begin{equation}
\label{gravpow}
{\cal P}_g(k) \, 
= \, 2 {{k^3} \over {2 \pi^2a^2}}\vert \mu_k \vert ^2\, .
\end{equation}
The two point function appearing in (\ref{gravpow}) is that of a free
canonically normalized massless scalar field multiplied by $2 / m_{\rm
Pl}^2$. The factor $2$ comes from the fact that gravitational waves
have two independent states of polarization.  In analogy to
(\ref{gravpow}), the power spectrum of the curvature fluctuation
${\cal R}$ is
\begin{equation}
\label{scalepow}
{\cal P}_{\cal R}(k) \, 
= \, {{k^3} \over {2 \pi^2}z^2}\vert v_k \vert ^2\, .
\end{equation}
This last quantity can be estimated very easily. From the fact that 
on scales larger than the Hubble radius the mode
functions are proportional to $a(\eta)$, we find
\begin{equation}
\label{final}
{\cal P}_{\cal R}(k) \simeq {{k^3} \over {2 \pi^2}} {1 \over 2k}
\frac{1}{a^2[\eta_{\rm H}(k)]} \, ,
\end{equation}
where $\eta_{\rm H}(k)$ is the conformal time of Hubble radius crossing for
the mode with comoving wavenumber $k$. Note that the second factor on
the r.h.s. of (\ref{final}) represents the vacuum normalization of the
wavefunction.

\section{Why Significant Effects of Trans-Planckian Physics
on CMB Anisotropies are Possible}

Let us now return to the main topic of this Letter, namely the
question of why significant effects of trans-Planckian (string)
physics on CMB anisotropies are possible. They key point is that in
most current models of inflation, the duration of the phase of
quasi-exponential expansion is so long that at the beginning of this
period, the time when the initial conditions for the fluctuations are
set, the physical wavelengths of modes responsible for the CMB
anisotropies are smaller than the Planck (string) scale. Let us
illustrate this point with a concrete example. In a single-field 
model of inflation, the number of e-folds is given by the formula
\begin{equation}
\label{efold}
N=-\frac{8\pi }{m_{\rm Pl}^2}\int _{\varphi _{\rm i}}
^{\varphi _{\rm e}} {\rm d}\varphi V(\varphi ) 
\biggl(\frac{{\rm d}V}{{\rm d}\varphi}\biggr)^{-1}\, ,
\end{equation}
where $\varphi _{\rm i}$ is the value of the scalar field at the
beginning of inflation and $\varphi _{\rm e}$ is the value at the end
of inflation, i.e., when $-\dot{H}/H^2=1$ (a dot means a derivative
with respect to cosmic time $t$). Let us consider the prototypical
model of inflation, namely chaotic inflation with a
scalar field potential which is given
by $V(\varphi )=(\lambda/4!)  \varphi ^4$ with $\lambda \simeq
10^{-15}$. In this case $\varphi _{\rm e}=m_{\rm Pl}/\sqrt{\pi }$.
The integral in Eq.~(\ref{efold}) can be easily performed and we get
\begin{equation}
N=\frac{\pi }{m_{\rm Pl}^2}\varphi _{\rm i}^2-1\, .
\end{equation}
In chaotic inflation the initial conditions are fixed according to the
rule $V(\varphi _{\rm i})\simeq m_{\rm Pl}^4$ which amounts to
$\varphi _{\rm i}\simeq (24/\lambda )^{1/4} m_{\rm Pl}$. As a
consequence, we deduce that $N\simeq 2\pi \sqrt{6/\lambda }-1\simeq
4.9\times 10^8$. This means that the Hubble radius today, $\ell _{\rm
H}=10^{61}\ell _{\rm Pl}$ ($h=0.5$), where $\ell _{\rm Pl}$ is the
Planck length, was equal to $\simeq e^{-10^8}\ell _{\rm Pl}\simeq
10^{-4.7\times 10^7}\ell _{\rm Pl}$ at the beginning of
inflation, i.e., very well below the Planck length indeed. 
On these scales, it is clear that the framework of standard 
quantum field theory described in the previous section and used 
in order to establish what the predictions of inflation are 
most likely breaks down.
\par

This simple example raises the following question: do the predictions
of inflation depend on physics on scales shorter than the Planck
length? In trying to answer this question we immediately face the
problem that the trans-Planckian physics is presently unknown and
that, as a consequence, it is impossible to study its impact on the
inflationary predictions. To circumvent this difficulty, one studies
the robustness of inflationary predictions to ad-hoc (reasonable)
changes in the standard quantum field theory framework supposed to
mimic the modifications caused by the actual theory of quantum
gravity. If the predictions are robust to some reasonable changes,
then it seems likely that they will be robust to the modifications
induced by the true theory of quantum gravity. On the other hand, if
the predictions are not robust, the knowledge of the exact theory is
required in order to predict exactly what the changes are. However,
one can still study what is the origin of these modifications and try
to use the currently available data to put constraints on the unknown
theory of quantum gravity.
\par

\begin{widetext}

\begin{figure}[htbp]
\includegraphics*[width=18cm, height=10cm, angle=0]{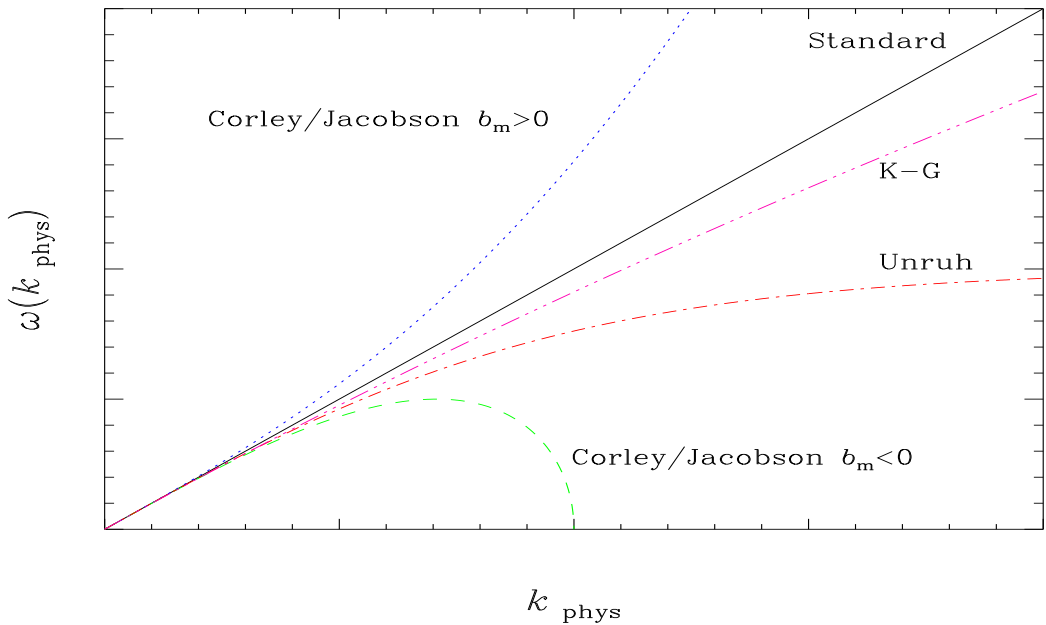}
\caption{Various dispersion relations considered in the literature (see 
Refs.~\cite{Unruh,CJ,KG}, and also Ref.~\cite{Mersini} for another 
dispersion relation not displayed on the plot). For $k_{\rm phys}
\ll k_{_{\rm C}}=k_{_{\rm Pl}}$, all the dispersion 
relations are linear, $\omega_{\rm phys}\simeq k_{\rm phys}$, 
which guarantees that the laws of physics on scales below the 
Planck scale are the standard ones. On the other hand on very 
small scales, for $k_{\rm phys}\gg k_{_{\rm C}}$, the dispersion 
relations deviate from the standard one.}
\label{disp}
\end{figure}
\end{widetext}

Let us now describe what are the possible ad-hoc modifications that
can be considered. So far, four different possibilities have been
studied. The first one consists in replacing the standard dispersion
relation $\omega =k$ by some ad-hoc relations. This proposal was first
made in Refs.~\cite{Unruh,CJ} in the context of black hole physics and is
based on an analogy with condensed matter physics. It is known that
the dispersion relation starts departing from the linear relation
$\omega =k$ on scales of the order of the atomic separation: the mode
feels the granular nature of matter. In the same way, one can expect
the dispersion relation to change when the mode starts feeling the
discreteness of space-time on scales of the order of the Planck (string)
length. Some of the dispersion relations studied so far
in the literature  are displayed in Fig.~\ref{disp}. 
We will come back to this possibility below.
\par

The second proposal is to modify the standard commutation relations.
The modification envisaged in Refs.~\cite{Ve,GM,ACV,GKP} amounts to
choose
\begin{equation}
[\hat{x},\hat{p}]=i\hbar(1+\beta \hat{p}^2),
\end{equation}
which introduces a short distance cut-off. Its implications for the wave 
equation were studied in Ref.~\cite{kempf}. It has been argued that
this commutation relation could be a generic prediction of string
theory. It has been shown in Refs.~\cite{Chu,Easther,kempfN} that
significant effect on the power spectrum are possible in this case.
\par

The third proposal consists in assuming that the quantum state 
in which the cosmological perturbations are placed is no longer 
the vacuum but some excited state. In this case, it has been 
shown in Ref.~\cite{Hui} that a possible observable signature 
is the modification of the consistency relations
between the spectral indices of scalar and tensor fluctuations
in inflation. Proposals 2 and 3 are not necessarily unrelated. 
For example, in the work of Ref.~\cite{Easther}, the modified 
commutation relation leads to a state at Hubble radius crossing 
which is not the adiabatic vacuum.
\par

Very recently, a fourth possibility has been suggested in
Ref.~\cite{Kaloper} using general effective field theory
techniques. In this case, it has been demonstrated that
trans-Planckian effects on CMB anisotropies in an inflationary
Universe are suppressed by a factor of $(H/M)^2$, where $H$ is the
Hubble constant during inflation, and $M$ is the scale of
trans-Planckian physics. If we take into account that the value of $H$
during inflation is bounded from above due to the observational bounds
on the spectrum of gravitational waves \cite{Hbound}, then - in models
in which the string scale is close to the usual four-dimensional
Planck scale - this suppression factor render the signatures of this
kind of trans-Planckian physics far too small to be observed in the
near future. On this basis, the authors of Ref.~\cite{Kaloper}
conclude that short distance physics cannot be observed in the
CMB. However, this conclusion clearly rests on the type of modifications 
chosen by the authors and is not true in general.
\par

To demonstrate that {\it a priori} significant changes in the
inflationary predictions are possible, let us come back to the case
where the standard dispersion is modified.  The method is to replace
the linear dispersion relation $\omega _{_{\rm phys}}=k_{\rm phys}$ by
a non standard dispersion relation $\omega _{_{\rm phys}}=\omega
_{_{\bf phys}}(k)$. In the context of cosmology, it has been shown in
Ref.~\cite{MB} that this amounts to replacing $k^2$ appearing in
(\ref{scaleeq}) with $k_{\rm eff}^2(n,\eta )$ defined by
\begin{equation}
k^2 \, \rightarrow \, k_{\rm eff}^2(k,\eta ) \equiv 
a^2(\eta )\omega _{_{\rm phys}}^2\biggl[\frac{k}{a(\eta )}\biggr].
\end{equation}
For a fixed comoving mode, this implies that the dispersion relation
becomes time-dependent. Therefore, the equation of motion of the
quantity $v_k(\eta)$ takes the form
\begin{equation} 
\label{eom2}
v_k'' + \biggl[k_{\rm eff}^2(k,\eta ) - {{a''} 
\over a}\biggr]v_k \, = \, 0 \, .
\end{equation}
Let us remark that a more rigorous derivation of this 
equation, based on a variational principle, has been provided 
in Ref.~\cite{LLMU}, see also Refs.~\cite{jacobson,jacobson2}.
\par

The evolution of modes thus must be considered separately in three
phases, see Fig.~\ref{sketch}. In Phase I the wavelength is smaller
than the Planck scale, and trans-Planckian physics is expected to play
an important role. In Phase II, the wavelength is larger than the
Planck scale but smaller than the Hubble radius. In this phase,
trans-Planckian physics is expected to have a negligible effect (and
the work of \cite{Kaloper} makes this statement quantitative). Hence,
by the analysis in Section II, the wavefunction of fluctuations is
oscillating in this phase, 
\begin{equation}
\label{vsubH}
v_k \, = \, B_1\exp(-ik\eta )+B_2\exp(ik\eta )
\end{equation}
with constant coefficients $B_1$ and $B_2$. In the standard approach,
the initial conditions are fixed in this region and the usual choice
of the vacuum state leads to $B_1=1/\sqrt{2k}$, $B_2=0$ (see the
previous section). Phase III starts at the time $t_{\rm H}(k)$ when the
mode crosses the Hubble radius. During this phase, the wavefunction is
squeezed and is given by
\begin{equation}
\label{vsuperH}
v_k \, = \, C_1z(\eta )+C_2z(\eta )\int ^{\eta }\frac{{\rm d}\tau 
}{z^2(\tau )}\, .
\end{equation}
The source of trans-Planckian effects on observations studied in
\cite{MB} is the possible non-adiabatic evolution of the wavefunction
during Phase I. If this occurs, then it is possible that the
wavefunction of the fluctuation mode is not in its vacuum state when
it enters Phase II and, as a consequence, the coefficients $B_1$ and
$B_2$ are no longer given by the standard expressions above. In this
case, the wavefunction will not be in its vacuum state when it crosses
the Hubble radius, and the final spectrum will be different. In
general $B_1$ and $B_2$ are determined by the matching conditions
between phase I and II. If the dynamics is adiabatic
throughout (in particular if the $a''/a$ term is negligible), the WKB
approximation holds and the solution is always given by
\begin{equation} 
\label{WKBsol}
v_k (\eta )\, \simeq \, \frac{1}{\sqrt{2k_{\rm eff}(k,\eta )}}
\exp\biggl[-i\int _{\eta _{\rm i}}^{\eta }k_{\rm eff}{\rm d}\tau \biggr]
\, ,
\end{equation} 
where $\eta_i$ is some initial time. Therefore, if we start with
a positive frequency solution only and uses this solution, one finds
that no negative frequency solution appears. Deep in the region II where
$k_{\rm eff} \simeq k$ the solution becomes
\begin{equation}
v_k(\eta ) \simeq {1 \over {\sqrt{2k}}} \exp(-i \phi - i k \eta),
\end{equation}
i.e. the standard vacuum solution times a phase which will disappear
when we calculate the modulus. The phase $\phi $ is given by $\phi
\equiv \int _{\eta _{\rm i}}^{\eta _1}k_{\rm eff} {\rm d}\tau $, where
$\eta _1$ is the time at which $k_{\rm eff}\simeq k$. By focusing only on
trans-Planckian effects on the local vacuum wave function at the
time $t_{\rm H}(k)$, the authors of \cite{Kaloper} miss this important
potential source of trans-Planckian signals in the CMB.
\par

It is possible to give the conditions for violation of adiabaticity
and to quantify exactly the accuracy of the WKB approximation. Given
an equation of the form $\mu ''+\omega ^2\mu =0$ (in the present
context, one has $\omega ^2=k^2_{\rm eff}-a''/a$), the WKB
approximation is valid if the quantity $\vert Q/\omega ^2 \vert \ll
1$, where the quantity $Q$ is defined by the following expression
\begin{equation}
Q=\frac{3(\omega ')^2}{4\omega ^2}-\frac{\omega ''}{2\omega }.
\end{equation}
This standard criterion can be obtained in the following manner. The
WKB solution, $\mu _{\rm wkb}$, satisfies the equation $\mu _{\rm
wkb}''+\mu _{\rm wkb}(\omega ^2-Q)=0$ exactly. Therefore, one has $\mu
\simeq \mu _{\rm wkb}$ if $\vert Q/\omega ^2 \vert \ll 1$. To obtain 
a modification of the inflationary spectrum, it is sufficient to 
find a dispersion relation such that $\vert Q/\omega ^2 \vert \gg 1$ 
in phase I.
\par

Let us now present some concrete examples where a change in the
inflationary spectrum has been obtained. The first dispersion relation
for which these effects were found is
\begin{equation}
\label{disprel}
k_{\rm eff}^2(k,\eta ) = k^2 - k^2 \vert b_m \vert
\biggl[{{\ell_{_{\rm C}}} \over {\lambda(\eta)}} \biggr]^{2m}, 
\end{equation}
where $\lambda (\eta )=2\pi a(\eta )/k$ is the wavelength of a
mode. For this case, it was found that the spectral index is modified
and that superimposed oscillations appear. However, important
concerns regarding the previous conclusion can be raised. For example,
the dispersion relation (\ref{disprel}) used leads to complex
frequencies in the context of an inflationary model with a long period
of superluminal expansion. Furthermore, the initial conditions for the
Fourier modes of the fluctuation field have to be set in the region
where the evolution is non-adiabatic and the use of the usual vacuum
prescription can be questioned. For this reason, the previous 
example is not satisfactory~\cite{MB2}.
\par

\begin{widetext}

\begin{figure}[htbp]
\includegraphics*[width=18cm, height=10cm, angle=0]{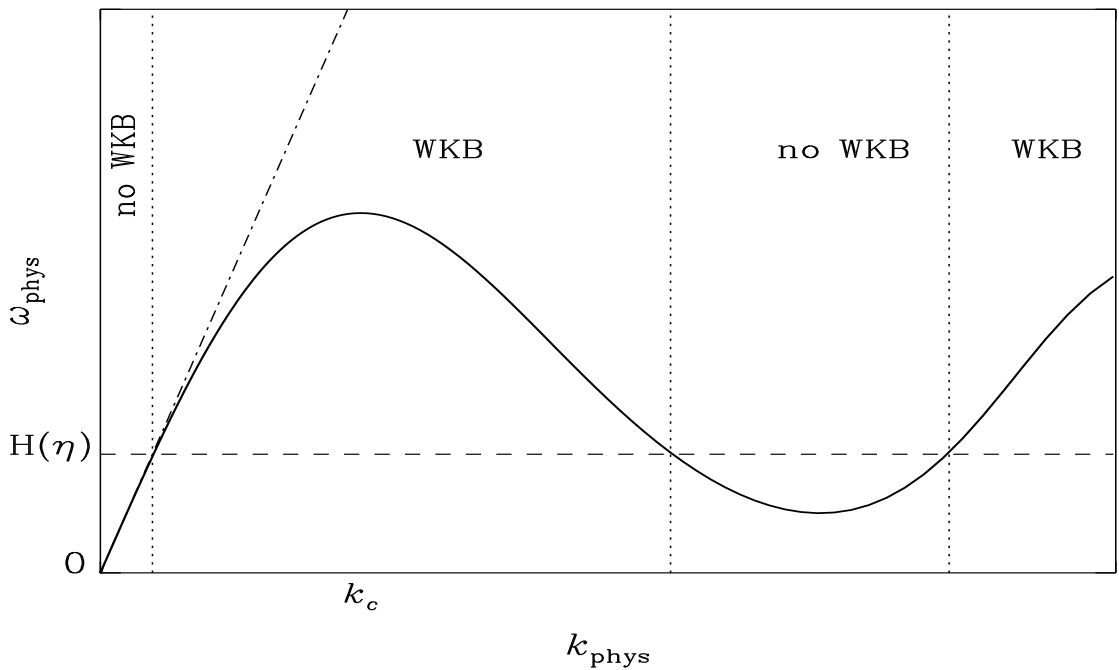}
\caption{Example of a dispersion relation where the WKB approximation 
can be violated during phase I.}
\label{LL}
\end{figure}
\end{widetext}

Examples where a modification can be obtained without the previous
difficulties can nevertheless be obtained. In Ref.~\cite{LLMU} such an
example has been explicitly constructed with the dispersion relation
\begin{equation}
\omega _{\rm phys}^2=k^2_{\rm phys}+2b_{11}k_{\rm phys}^4
-2b_{12}k_{\rm phys}^6.
\end{equation}
The coefficients $b_{11}$ and $b_{12}$ can be chosen 
such the dispersion relation has a maximum around $k_{_{\rm C}}$ 
and a minimum at a scale smaller than $k_{_{\rm C}}$. This minimum 
can be chosen to be smaller than the Hubble parameter during phase I. 
This is the case in Fig.~\ref{LL}.
In this example, $\omega _{\rm phys}$ always remains real, the initial
conditions can be fixed in a region where the WKB approximation is
valid (as a consequence, the initial state can be chosen as the minimal
energy state~\cite{Brown78}) and where the mode function $v_k$
oscillates. Since there is a phase during which the WKB approximation
is not valid, the final spectrum is modified. It has been calculated
explicitly in Ref.~\cite{LLMU}. The difficulties described previously
can also be avoided by considering the evolution of fluctuations in a
bouncing cosmology in which the initial conditions can be set in the
asymptotically flat past, and focusing on modes for which the
frequency never becomes complex \cite{BJM}. For all the examples where
everything can be done consistently, the coefficients $B_1(k)$ and
$B_2(k)$, see Eq.~(\ref{vsubH}), in phase II are found to be of the
form
\begin{eqnarray}
\label{Bogo1}
B_1(k) &=& \frac{1}{\sqrt{2k}}[1+\epsilon \xi _1(k)
+{\cal O}(\epsilon ^2)]\, , 
\\
\label{Bogo2}
B_2(k) &=& \epsilon \xi _2(k)+{\cal O}(\epsilon ^2)\, ,
\end{eqnarray}
where the functions $\xi _1(k)$ and $\xi _2(k)$ have been 
explicitly calculated in Refs.~\cite{LLMU,BJM}. In the previous 
relations, $\epsilon $ is a small parameter which is basically 
the time that the mode has spent in the region where the WKB 
approximation is violated. The advantage of expanding the 
two  coefficients in the parameter $\epsilon $ is that general 
equations can be obtained. But this does not mean that the 
correction has to be proportional to a small parameter and 
large non perturbative effects can also be obtained if the time 
spent by the mode in the region where WKB is not valid is 
large.
\begin{widetext}

\begin{figure}[htbp]
\includegraphics*[width=18cm, height=10cm, angle=0]{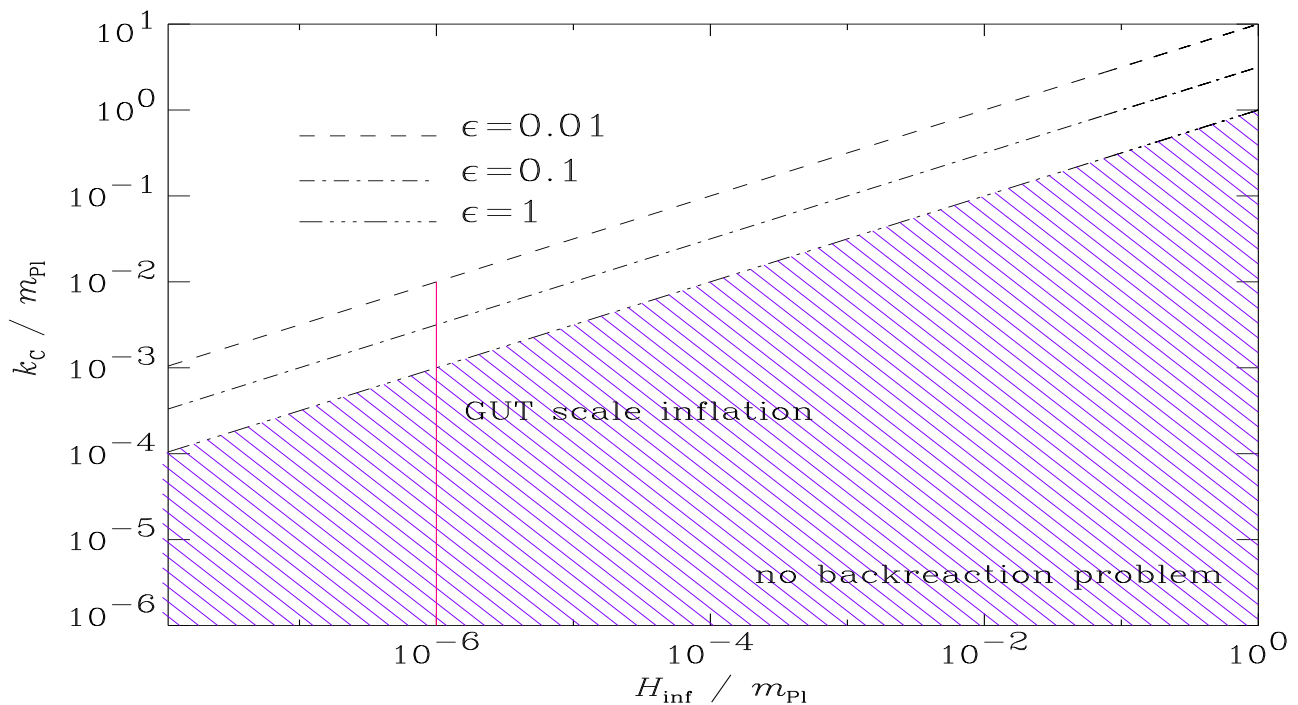}
\caption{Region in the $(k_{_{\rm C}}, H_{\rm inf})$ plan where 
a correction of order $\epsilon $ is expected with a back-reaction 
problem. The shaded region indicates the region obtained 
for $\epsilon =0.01$. For larger $epsilon $ the shaded region should 
be extended up to the corresponding straight line indicated on the 
figure.}
\label{backreac}
\end{figure}
\end{widetext}

In this case, another concern \cite{Tanaka,Starob} is that the
excitations produced during Phase I might have an important
back-reaction effect. This is because an excited state leads 
to an energy density that could be larger that the vacuum energy 
density causing inflation. The energy density due to trans-Planckian 
effect is~\cite{LLMU}
\begin{equation}
\rho  \simeq {\cal O}(\epsilon ^2k_{_{\rm C}}^4).
\end{equation}
Therefore, there is no back-reaction problem if $\rho <m_{\rm
Pl}^2H_{\rm inf}^2$. Obviously, the smallest $\epsilon $ is, the less
severe the back-reaction problem is but, at the same time, the less
important the modification of the spectrum is. The important point is
that, by playing with the quantities $H_{\rm inf}$ and $k_{_{\rm C}}$, a
window where the correction is significant and where there is no
back-reaction can be found. These  results are summarized in 
Fig.~\ref{backreac}.

In the light of the work of Refs.~\cite{Easther,kempfN}, another way
to think about the coefficients $B_1(k)$ and $B_2(k)$, see
Eqs.~(\ref{Bogo1}) and (\ref{Bogo2}), is as phenomenological parameters
which describe the effects of short distance physics. Trans-Planckian
(stringy) physics can lead to non-standard values of these
coefficients at the earliest time when the fluctuation modes can be
described by the usual actions for linearized gravitational
fluctuations discussed in Section II, and the work of
Ref.~\cite{Easther} presents a concrete model where values of $B_1(k)$
and $B_2(k)$ arise which differ from the standard values enough to
produce measureable effects but for which the back-reaction of the
non-vacuum state at Hubble radius crossing is negligible (We thank 
Brian Greene for communications on this point).

\section{Discussion and Conclusion}
 
Based on a review of the theory of cosmological fluctuations as
applied to inflationary cosmology we have discussed the main sources
of expected trans-Planckian (stringy) signatures on the spectrum of
CMB anisotropies. One important potential source~\cite{MB} is the fact
that the evolution of the fluctuation modes can be non-adiabatic when
their wavelength is smaller than the Planck (string) scale. This leads
to an excited state of the fluctuation modes at the time when the mode
crosses the Hubble radius at $t_{\rm H}(k)$. Another way in which
trans-Planckian effects can lead to an excited state of the
fluctuation modes at $t_{\rm H}(k)$ is that new
physics~\cite{Easther,kempfN} will generate non-trivial Bogoliubov
coefficients $B_1(k)$ and $B_2(k)$, see Eqs.~(\ref{Bogo1}) and
(\ref{Bogo2}), at the earliest time that the modes can be adequately
approximated by the usual actions for linearized gravitational
fluctuations.

The recent paper \cite{Kaloper} does not address this issue. It
focuses on the computation of the trans-Planckian (stringy) corrections
to the fluctuation amplitude in the local vacuum state at $t_{\rm H}(k)$,
and comes to the correct conclusion that these corrections are usually
negligible. If one were to try to use the effective field theory
techniques used in \cite{Kaloper} to address the evolution of the
modes in Phase I, one would not be able to integrate over degrees
of freedom with frequency larger than $H$. Since the frequencies of
the modes in Phase I are trans-Planckian, one is not allowed to
integrate out any sub-Planckian modes. From this
point of view one would reach the conclusion that one is
not in the regime of applicability of effective field theory,
and that therefore the corrections to results obtained (like
the standard results of inflation on the spectrum of fluctuations)
should be expected to be of order unity or larger.

We hope that the review of the theory of cosmological perturbations
presented in Section II will be of use to physicists not actively
working on cosmological perturbations, and that it will dispel
the myth that fluctuations are generated at Hubble radius crossing.
Note also that in the modern version of the theory of cosmological
perturbations presented here, there is no need for ad hoc ultraviolet
subtractions, since the analysis is done consistently in momentum
space in linear perturbation theory. A final caveat, however, is
that if it were to turn out that inflation is the result of some
highly nonlinear theory at a scale much smaller than the
Planck scale (see e.g. the model of \cite{BZ}), then
the theory of cosmological perturbations presented in Section II
would not apply on scales smaller than the Hubble radius. One
would then need to calculate in terms of an effective field theory,
and by the analysis of \cite{Kaloper} one would not expect any
deviations from the expected spectrum of scale-invariant
fluctuations due to trans-Planckian physics.

\vspace{0.5cm} 
\centerline{\bf Acknowledgments}
\vspace{0.2cm}

We would like to thank Brian Green and Paul Windey for comments and
careful reading of the manuscript and the authors of
Ref.~\cite{Kaloper} for sending us a draft of their paper before
publication and for useful comments on this manuscript. We acknowledge
support from the BROWN-CNRS University Accord and are grateful to Herb
Fried for his efforts to secure this Accord. R.~B.  wishes to thank
the CERN Theory Division for hospitality and support during the time
when this paper was written.  The research was also supported in part
(at Brown University) by the U.S. Department of Energy under Contract
DE-FG02-91ER40688, TASK A.

\end{document}